\documentclass[superscriptaddress,twocolumn,showpacs,prl,floatfix]{revtex4}

\usepackage{color}
\usepackage{tabularx}
\usepackage{epsfig}
\usepackage{amsmath}
\usepackage{graphicx}

\begin{document}

\title{The Ising ferromagnet in dimension five : link and spin overlaps}

\author{P. H.~Lundow}
\affiliation {Department of Theoretical Physics, Kungliga Tekniska h\"ogskolan, SE-106 91 Stockholm, Sweden}

\author{I. A.~Campbell}
\affiliation{Laboratoire Charles Coulomb,
  Universit\'e Montpellier II, 34095 Montpellier, France}

\begin{abstract}
In the simple [hyper]cubic five dimension near neighbor interaction
Ising ferromagnet, extensive simulation measurements are made of the
link overlap and the spin overlap distributions. These "two replica"
measurements are standard in the Spin Glass context but are not
usually recorded in ferromagnet simulations.  The moments and moment
ratios of these distributions (the variance, the kurtosis and the
skewness) show clear critical behaviors at the known ordering
temperature of the ferromagnet. Analogous overlap effects can be
expected quite generally in Ising ferromagnets in any dimension. The
link overlap results in particular, with peaks at criticality in the
kurtosis and the skewness, also have implications for Spin Glasses.

\end{abstract}


\maketitle

\section{Introduction}
As is well known, the upper critical dimension (ucd) for Ising
ferromagnets is four.  Here we show results of simulations for an
Ising ferromagnet with near neighbor interactions, on a simple
[hyper]cubic lattice with periodic boundary conditions in dimension
five, the next dimension up.  The Hamiltonian is as usual
\begin{equation}
  \mathcal{H}= - J\sum_{ij}S_{i}S_{j}
  \label{ham}
\end{equation}
with spins $i$ and $j$ near neighbors. We will take $J=1$ and will
quote inverse temperatures $\beta = 1/T$.

All the principal properties of this system are well known. Recent
consistent and precise estimates of the inverse ordering temperature
$\beta_c$ are $0.11391$ \cite{jones:05}, 0.113925(12)
\cite{berche:08}, $0.1139139(5)$ \cite{lundow:11} from simulations,
and 0.113920(1)\cite{butera:12} from High Temperature Series Expansion
(HTSE). We will use $\beta_c=0.113915(1)$ as a compromise estimate.
The susceptibility critical exponent and the effective correlation
length exponent take the exact mean field values $\gamma=1$ and $\nu =
1/2$, with a leading correction to scaling exponent $\theta=1/2$
\cite{guttmann:81}. In the periodic boundary condition geometry, above
the ucd the "effective length" is $L_{\mathrm{eff}}=AL^{d/4}$ where
$A$ is a non-universal constant \cite{brezin:82,jones:05}.

We analyse simulation data for the "two replica" observables link
overlap and spin overlap, familiar in the Ising Spin Glass (ISG)
context.  It might seem curious to apply techniques developed for
complex systems to the much simpler ferromagnet, particularly above
the ucd, even though the observables can be defined in exactly the
same way in a ferromagnet as in an ISG. Properties of the spin overlap
at and beyond $\beta_c$ have already been studied in the $3$d Ising
ferromagnet \cite{berg:02}.


However, just because all the major parameters are well known it is
convenient to use the $5$d ferromagnet as a testbed for studying the
critical behavior of various moments or moment ratios (the variance,
the kurtosis and the skewness) of both overlap distributions. The
results should be generalizable {\it mutatis mutandis} to all Ising
ferromagnets. The final aim is to establish the ground rules for the
properties related to the link overlap near criticality in order to
apply a similar methodology to complex systems, in particular to ISGs.

The link overlap \cite{caracciolo:90}, like the more familiar spin
overlap, is an important parameter in ISG numerical simulations.  In
both cases two replicas (copies) $A$ and $B$ of the same physical
system, i.e. with identical sets of interactions between spins, are
first generated and equilibrated; updating is then continued and the
"overlaps" between the two replicas are recorded over long time
intervals. The spin overlap at any instant $t$ corresponds to the
fraction $q(t)$ of spins $S_{i}$ in $A$ and $B$ having the same
orientation ($S_{i}^{A}$ and $S_{i}^{B}$ both up or both down), and
the normalized overall distribution over time is written $P(q)$. The
link overlap corresponds to the fraction $q_{\ell}(t)$ of links (or
bonds or edges) $ij$ between spins which are either both satisfied or
both dissatisfied in the two replicas; the normalized overall
distribution over time is written $Q(q_{\ell})$.  The explicit
definitions are
\begin{equation}
  q(t)=\frac{1}{N}\,\sum_{i=1}^{N} S_{i}^{A}(t)S_{i}^{B}(t)
  \label{qtdef}
\end{equation}
and
\begin{equation}
  q_{\ell}(t)=\frac{1}{N_{l}}\sum_{ij}S_{i}^{A}(t)S_{j}^{A}(t)S_{i}^{B}(t)S_{j}^{B}(t)
  \label{qltdef}
\end{equation}
where $N$ is the number of spins in the system and $N_{l}$ the number
of links; spins $i$ and $j$ are linked, as denoted by $ij$.  We will
indicate means taken over time by $\langle\cdots\rangle$.

Equilibration and measurement runs were performed by standard heat
bath updating on sites selected at random.  The spin systems started
with a random configuration, i.e. at infinite temperature, and were
gradually cooled and equilibrated until they reached their designated
temperatures, where they received a longer equilibration.  For
example, for $L=10$ the systems at $\beta=0.114$ saw roughly $10^6$
sweeps before any measurements took place and the smaller systems at
least $10^7$ sweeps. Several sweeps were made between measurements and
with a flip rate of about $27\%$ near $\beta_c$ which means at least
four sweeps between measurements. For $L=10$ about $10^7$ measurements
were recorded at each temperature near $\beta_c$ and considerably
more for the smaller systems.

\section{Link overlap}

For any standard near neighbor Ising ferromagnet with all interactions
identical and with periodic boundary conditions, there is a simple
rule for the mean link overlap in equilibrium $\langle
q_{\ell}(\beta,L)\rangle$.  If $p_{s}(\beta,L)$ is the probability
averaged over time that any given bond is satisfied, then by
definition the mean energy per bond is
\begin{equation}
  |U(\beta,L)| \equiv 1-2p_{s}(\beta,L)
\end{equation}
Because all bonds are equivalent
\begin{equation}
  \langle q_{\ell}\rangle(\beta,L) = p_{s}^2 + (1-p_{s})^2 - 2p_{s}(1-p_{s}) \equiv  U(\beta,L)^2.
\end{equation}
This rule is exact at all temperatures (we have checked this
numerically), so it would appear at first glance that link overlap
measurements present no interest in a simple ferromagnet as they
contain no more information than the energy. However, the moments of
the link overlap distribution reflect the structure of the temporary
spin clusters which build up in the paramagnetic state before
$\beta_c$, and the domain structure in the ferromagnetic state beyond
$\beta_c$. Thus if at some instant $t$ a cluster of parallel spins
exists in replica $A$ and a similar cluster in the same part of space
exists in replica $B$, then the instantaneous $q_{\ell}(t)$ will be
significantly higher than the time average $\langle
q_{\ell}(t)\rangle$. The width of the overall distribution
$Q(q_{\ell})$ increases rapidly on the approach to $\beta_c$ and we
find phenomenologically that, as a consequence of the repeated
occurrence of the cluster situation, around the critical temperature
the distributions do not remain simple Gaussians but develop excess
kurtosis and skewness, even though to the naked eye these deviations
from pure Gaussian distributions are not obvious; for instance no
secondary peaks appear in the distributions.

We exhibit in Figures 1 to 4 data at sizes $L=4,6, 8$ and $10$ for the
Q-variance
\begin{equation}
  Q_{\mathrm{var}}(\beta,L) = \left\langle\left(q_{\ell}-\langle q_{\ell}\rangle\right)^2\right\rangle,
  \label{Qvar}
\end{equation}
the Q-kurtosis
\begin{equation}
  Q_{k}(\beta,L) =
  \frac{
    \left\langle\left(q_{\ell}-\langle q_{\ell}\rangle\right)^4\right\rangle
  }{
    \left\langle\left(q_{\ell}-\langle q_{\ell}\rangle\right)^2\right\rangle^2
  }
  \label{Qkurt}
\end{equation}
and the Q-skewness
\begin{equation}
  Q_{s}(\beta,L) =
  \frac{
    \left\langle\left(q_{\ell}-\langle q_{\ell}\rangle\right)^3\right\rangle
  }{
    \left\langle\left(q_{\ell}-\langle q_{\ell}\rangle\right)^2\right\rangle^{3/2}
  }
  \label{Qskew}
\end{equation}

\begin{figure}
  \includegraphics[width=3.5in]{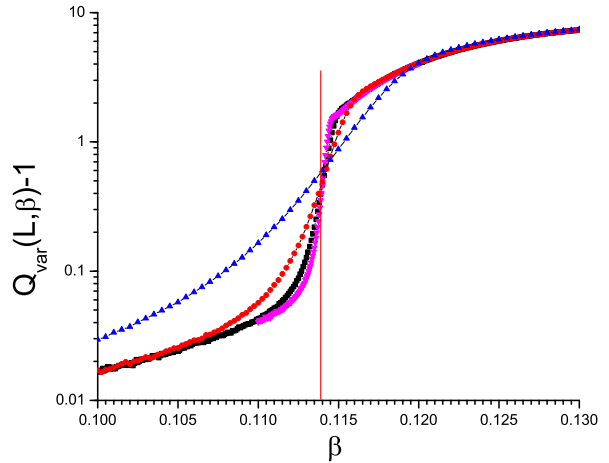}
  \caption{(Color online) The Q-variance, Eq.~\ref{Qvar}, as a
    function of size and inverse temperature for the $5$d near
    neighbor ferromagnet. In this and all the following figures the
    convention for indicating size is : $L=4$, blue triangles; $L=6$,
    red circles; $L=8$, black squares; $L=10$, pink inverted
    triangles.  In this and following figures errors are smaller than
    the size of the points unless stated otherwise. The red vertical
    line indicates the inverse ordering temperature $\beta_{c}=
    0.113915$.}\protect\label{fig:1}
\end{figure}

\begin{figure}
  \includegraphics[width=3.5in]{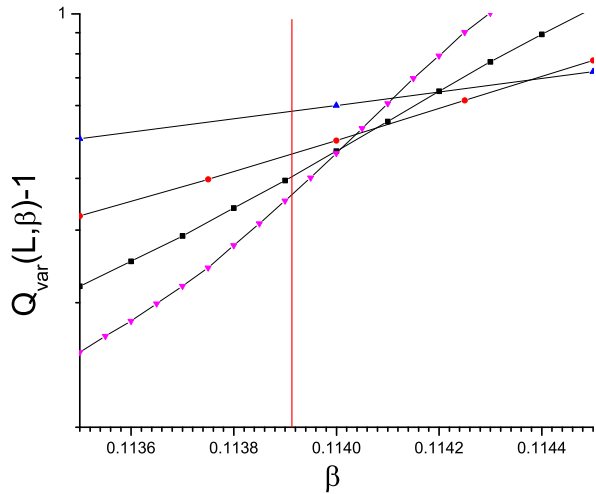}
  \caption{(Color online) The Q-variance as in Fig.~\ref{fig:1}, in
    the region of the inverse ordering temperature $\beta_c$. Sizes
    coded as in Fig.~\ref{fig:1}.  
  }\protect\label{fig:2}
\end{figure}

The three Q moments and moment ratios follow the standard definitions
for the moments of a distribution.  For the Q-variance we plot
$\log(Q_{\mathrm{var}}(\beta,L)-1)$ so as to display the entire range of
data.  Fig.~\ref{fig:1} shows the behavior of the Q-variance from high
temperature through $\beta_c$ to $\beta = 0.13$, well into the ordered
state. Fig.~\ref{fig:2} shows the same data in the region near
$\beta_c$.


It can be seen that the Q-variance has clear critical behavior. Just
as for standard "phenomenological couplings" such as the Binder
cumulant or the correlation length ratio $\xi(\beta,L)/L^{5/4}$ (in
the $5$d case \cite{jones:05}), it is size independent at $\beta_c$ to
within a finite size correction.  As the Q-variance is a
phenomenological coupling in this particular system, it can be
expected to have a similar form as a function of temperature in any
ferromagnet, with the appropriate finite size correction
exponent. This makes the Q-variance a supplementary phenomenological
coupling for ferromagnets in general. As $q_{\ell}$ is a near-neighbor
measurement like the energy, the distribution $Q(q_{\ell})$ tends to
equilibrate faster on annealing than a parameter such as the
correlation length $\xi$.
\begin{figure}
  \includegraphics[width=3.5in]{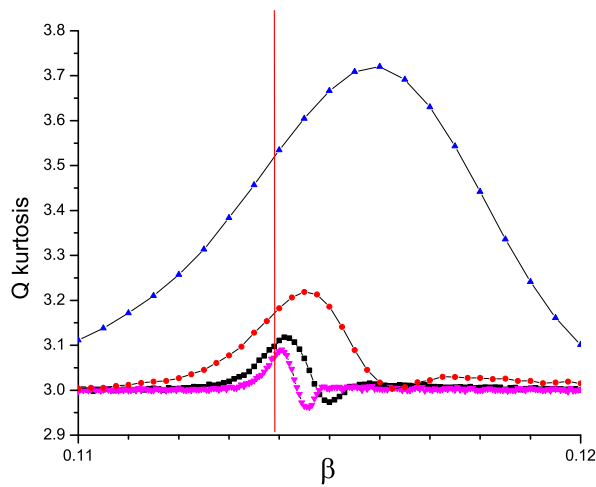}
  \caption{(Color online) The Q-kurtosis, Eq.~\ref{Qkurt}, for the
    $5$d near neighbor ferromagnet as a function of size and inverse
    temperature. Sizes coded as in
    Fig.~\ref{fig:1}. 
  }\protect\label{fig:3}
\end{figure}
\begin{figure}
  \includegraphics[width=3.5in]{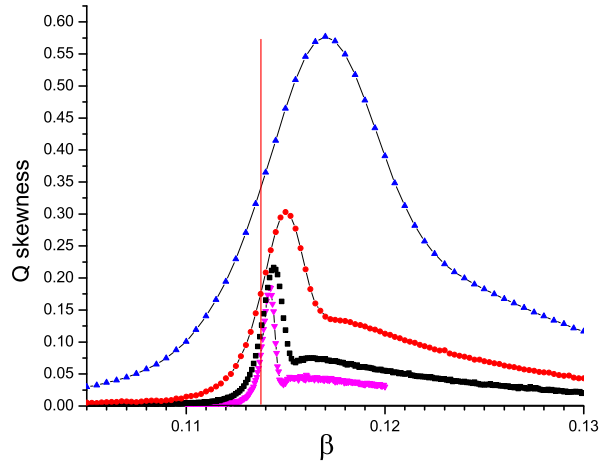}
  \caption{(Color online) The Q-skewness, Eq.~\ref{Qskew}, for the
    $5$d near neighbor ferromagnet as a function of size and inverse
    temperature. Sizes coded as in
    Fig.~\ref{fig:1}.
  }\protect\label{fig:4}
\end{figure}

The Q-kurtosis has a more unusual form, Fig.~\ref{fig:3}. At
temperatures well above or well below the critical temperature it
takes up the Gaussian value $Q_{k}(\beta)= 3$, but near criticality
there is an excess Q-kurtosis peak corresponding to a "fat tailed"
form of the link overlap distribution. With increasing $L$ the width
and strength of the peak decrease and the peak position
$\beta_{\mathrm{max}}(Q_{k})$ approaches $\beta_c$. In the present
$5$d ferromagnet case the form of the temperature dependence evolves
with $L$; with increasing $L$ from a simple peak it tends to peak plus
dip.

The Q-skewness, Fig.~\ref{fig:4}, resembles the Q-kurtosis plot. The
Q-skewness starts at $0$ (a symmetric distribution) at $\beta=0$ and
then develops a strong positive peak as a function of $\beta$ in the
region of $\beta_c$ (so a distribution $Q(q_{\ell})$ tilted towards
high $q_{\ell}$). Again the width and the strength of the peak
decrease with increasing $L$. There is a weak indication of the
beginning of a dip beyond the peak.  The Q-kurtosis and Q-skewness can
be expected to show qualitatively the same critical peak form in any
ferromagnet.

For the moment these observations are essentially phenomenological; it
would be of interest to go beyond the argument given above in terms of
correlated clusters of spins so as to obtain a full quantitative
explanation for the details of the critical behavior of the
$Q(q_{\ell})$ distribution and its moments in finite $L$ samples. Link
overlap moment peaks in ISGs resemble these ferromagnet results
\cite{lundow:12} implying that the peak structure is a very general
qualitative form of the behavior of link overlap distributions at an
Ising magnet critical point.


The link overlap can be defined for vector spins \cite{katzgraber:02} by
\begin{equation}
  q_{\ell} = \frac{1}{N_l}\sum_{ij}[({\bf S}^{A}_{i} \cdot {\bf S}^{A}_{j})({\bf S}^{B}_{i} \cdot {\bf S}^{B}_{j})]
  \label{qlvec}
\end{equation}
which is invariant under global symmetry operations; the same link
overlap critical properties as seen in Ising systems may well exist in
$XY$ and Heisenberg magnets also.

\section{Spin overlap}

As for the link overlap, one can also define various moments and moment ratios of the
spin overlap distribution such as the P-variance
\begin{equation}
  P_{\mathrm{var}}(\beta,L) = \langle q^2\rangle
  \label{Pvar}
\end{equation}
and the P-kurtosis
\begin{equation}
  P_{k}(\beta,L) = \langle q^4\rangle\big/\langle q^2\rangle^2
  \label{Pkurt}
\end{equation}
which is simply related to the Binder-like P-cumulant $P_{b}(\beta,L)
= \left(3-P_{k}(\beta,L)\right)/2$.

\begin{figure}
  \includegraphics[width=3.5in]{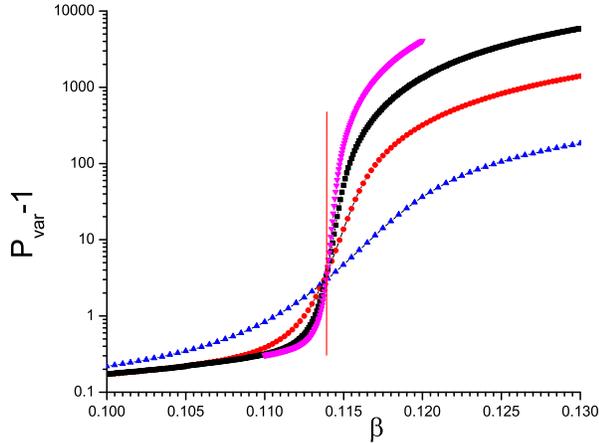}
  \caption{(Color online) The P-variance Eq.~\ref{Pvar} for the $5$d
    near neighbor ferromagnet as a function of size and inverse
    temperature. Sizes coded as in Fig.~\ref{fig:1}.
  }\protect\label{fig:5}
\end{figure}
\begin{figure}
  \includegraphics[width=3.5in]{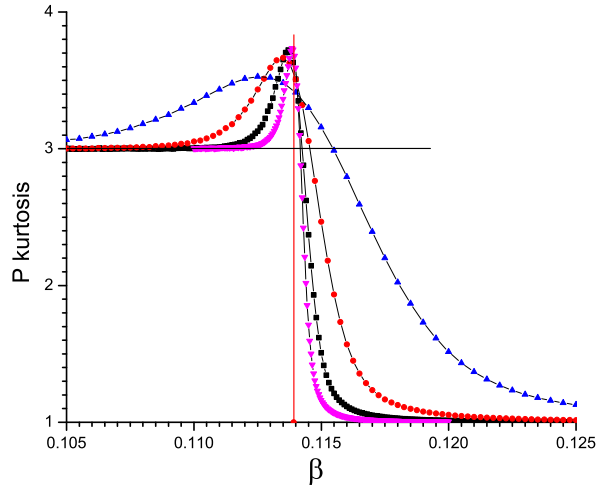}
  \caption{(Color online) The P-kurtosis Eq.~\ref{Pkurt} for the $5$d
    near neighbor ferromagnet as a function of size and inverse
    temperature. Sizes coded as in
    Fig.~\ref{fig:1}.
  }\protect\label{fig:6}
\end{figure}

The P-variance in the ferromagnet has the phenomenological coupling
form, Fig.~\ref{fig:5}; at $\beta_c$ the P-variance tends to an $L$
independent value with a finite size correction.  As the temperature
goes to zero the P-variance will tend to $L^d$.  The P-kurtosis has a
different phenomenological coupling form, a peak before a sharp drop
to $1$, corresponding to the $P(q)$ distribution becoming "fat tailed"
before and at $\beta_c$ before taking on a two-peak structure in the
ordered state~\cite{berg:02}. This is in contrast to the magnetization
M-kurtosis (usually expressed as the Binder cumulant) in ferromagnets
or the standard P-kurtosis in ISGs which both drop regularly from the
Gaussian value $3$ towards the two peak value $1$ with increasing
order. However, it can be noted that the temperature variation of the
kurtosis for the chiral order parameter in Heisenberg spin glasses has
the same general form as the present ferromagnetic P-kurtosis, the
distribution becoming fat tailed above the ordering temperature
\cite{hukushima:05}.

As the distributions $P(q)$ in equilibrium are by definition
symmetrical about $q=0$, the P-skewness is always zero.  However, for
the one-sided distribution of the absolute value of $|q|$, other
parameters can be defined, in particular the absolute P-kurtosis
\begin{equation}
  P_{|q|,k}(\beta,L) =
  \frac{
    \left\langle\left(|q|-\langle |q|\rangle\right)^4\right\rangle
  }{
    \left\langle\left(|q|-\langle |q|\rangle\right)^2\right\rangle^2
  }
  \label{Pabskurt}
\end{equation}
and the absolute P-skewness
\begin{equation}
  P_{|q|,s}(\beta,L) =
  \frac{
    \left\langle\left(|q|-\langle |q|\rangle\right)^3\right\rangle
  }{
    \left\langle\left(|q|-\langle |q|\rangle\right)^2\right\rangle^{3/2}
  }
  \label{Pabssk}
\end{equation}

\begin{figure}
  \includegraphics[width=3.5in]{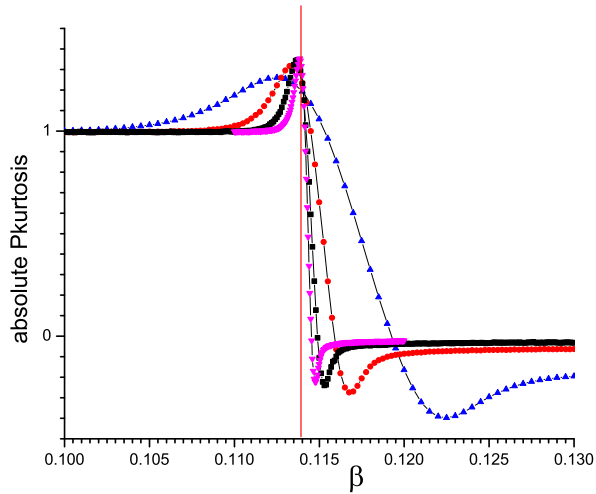}
  \caption{(Color online) The absolute P-kurtosis $P_{|q|,k}$,
    Eq.~\ref{Pabskurt}, for the $5$d near neighbor ferromagnet as a
    function of size and inverse temperature. Sizes coded as in
    Fig.~\ref{fig:1}.
  }\protect\label{fig:7}
\end{figure}

\begin{figure}
  \includegraphics[width=3.5in]{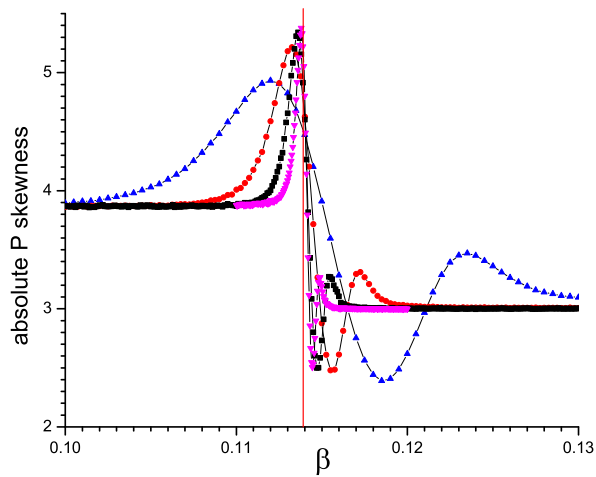}
  \caption{(Color online) The absolute P-skewness $P_{|q|,s}$,
    Eq.~\ref{Pabssk}, for the $5$d near neighbor ferromagnet as a
    function of size and inverse temperature. Sizes coded as in
    Fig.~\ref{fig:1}.
  }\protect\label{fig:8}
\end{figure}

The absolute P-kurtosis and the absolute P-skewness in the ferromagnet
have rather complex phenomenological coupling temperature dependence
patterns, with very weak finite size corrections at $\beta_c$,
Figs.~\ref{fig:7}, and \ref{fig:8}. If the weak finite size correction
is a general property for these parameters, it could be usefully
exploited so as to obtain high precision estimates of ordering
temperatures in systems where these temperatures are not well known.

The spin overlap properties do not transport from the ferromagnet into
ISG systems in the same manner as the link overlap properties do
because the P-variance takes on a different status : in an ISG $q^2$
becomes the order parameter.


\section{Conclusion}

The standard near neighbor interaction Ising ferromagnet on a simple
[hyper]cubic lattice in dimension five has been used as a test case in
order to demonstrate the critical form of the temperature variations
of observables related to the link overlap and the spin overlap,
parameters developed in the ISG context and not generally recorded in
ferromagnet simulations. The moments of the link and spin overlap
distributions $Q(q_\ell)$ and $P(q)$ show a rich variety of
temperature variations, with specific critical behaviors. The
temperature dependence of the link overlap kurtosis and the link
overlap skewness show peaks at criticality which are "evanescent" in
the sense that they will disappear in the the large size thermodynamic
limit. The present results validate the assumption that these
observables show true critical temperature dependencies.  A temporary
cluster phenomenon is proposed as determining the evolution of the
link overlap distributions. If correct, this mechanism is quite
general, so we expect the critical peaks in the Q-kurtosis and the
Q-skewness to be present in the entire class of Ising ferromagnets,
not only those in dimensions above the ucd, and plausibly in vector
spin ferromagnets also. Beyond the class of ferromagnets, it can be
noted that link overlap Q-kurtosis and Q-skewness critical peaks have
also been observed \cite{lundow:12} in Ising Spin Glasses.

\end{document}